# 3D XY Universality and Nonlinear magnetic susceptibility in a kagome ice compound

Kan Zhao[1, 2*], Hao Deng[3, 4, 5*], Hua Chen[6, 7*], Nvsen Ma[1*], Noah Oefele[2], Jiesen Guo[1], Xueling Cui[1], Chen Tang[1], Matthias J. Gutmann[8], Thomas Mueller[9], Yixi Su[9], Vladimir Hutanu[5], Changqing Jin[10, 11] and Philipp Gegenwart[2*]

[1]School of Physics, Beihang University, Beijing 100191, China

[2]Experimentalphysik VI, Center for Electronic Correlations and Magnetism, University of Augsburg, 86159 Augsburg, Germany.

[3]School of Physical Science and Technology, ShanghaiTech University, Shanghai 201210, China

[4]ShanghaiTech Laboratory for Topological Physics, ShanghaiTech University, Shanghai 201210, China

[5]Institute of Crystallography, RWTH Aachen University and Jülich Centre for Neutron Science (JCNS) at Heinz Maier-Leibnitz Zentrum (MLZ), Garching, Germany.

[6]Department of Physics, Colorado State University, Fort Collins, CO 80523, USA.

[7]School of Materials Science and Engineering, Colorado State University, Fort Collins, CO 80523, USA.

[8]ISIS Facility, Rutherford Appleton Laboratory, Chilton, Didcot OX11 0QX, UK

[9]Jülich Center for Neutron Science JCNS-MLZ, Forshungszentrum Jülich GmbH, Outstation at MLZ, D-85747 Garching, Germany

[10]Beijing National Laboratory for Condensed Matter Physics, Institute of Physics, Chinese Academy of Sciences, Beijing 100190, China

[11]School of Physical Sciences, University of Chinese Academy of Sciences, Beijing 100190, China

*Corresponding author email: kan_zhao@buaa.edu.cn, denghao@shanghaitech.edu.cn, huachen@colostate.edu, nvsenma@buaa.edu.cn, and philipp.gegenwart@physik.uni-augsburg.de

**Abstract**

Kagome spin ice is an intriguing class of spin systems constituted by in-plane *Ising* spins with ferromagnetic interaction residing on the kagome lattice, theoretically predicted to host a plethora of magnetic transitions and excitations. In particular, different variants of kagome spin ice models can exhibit different sequences of symmetry breaking upon cooling from the paramagnetic to the fully ordered ground state. Recently, it has been demonstrated that the frustrated intermetallic HoAgGe stands as a faithful solid-state realization of kagome spin ice. Here we use single crystal neutron diffuse scattering to map the spin ordering of HoAgGe at various temperatures more accurately and surprisingly find that the ordering sequence appears to be different from previously known scenarios: From the paramagnetic state, the system first enters a partially ordered state with fluctuating magnetic charges, in contrast to a charge-ordered paramagnetic phase before reaching the fully ordered state. Through state-of-the-art Monte Carlo simulations and scaling analyses using a quasi-2D model for the distorted Kagome spin ice in HoAgGe, we elucidate a single three-dimensional (3D) XY phase transition into the ground state with broken time-reversal symmetry (TRS). However, the 3D XY transition has a long crossover tail before the fluctuating magnetic charges fully order. More interestingly, we find both experimentally and theoretically that the TRS breaking phase of HoAgGe features an unusual, hysteretic response: In spite of their vanishing magnetization, the two time-reversal partners are distinguished and selected by a nonlinear magnetic susceptibility tied to the kagome ice rule. Our discovery not only unveils a new symmetry breaking hierarchy of kagome spin ice, but also demonstrates the potential of TRS-breaking frustrated spin systems for information technology applications.

Geometrical frustration in spin systems can result in exotic phases of matter (1-8). In two dimensions (2D), in-plane *Ising* spins with dominant nearest-neighbor ferromagnetic coupling on the kagome lattice, known as the kagome spin ice, has been a rich playground for unconventional critical behavior due to frustration (9-27). If there is only the nearest-neighbor ferromagnetic coupling $J_1$, the ground state of the model has an extensive entropy of 0.501$k_B$ per spin. This degeneracy can be removed with the help of either a 2nd-neighbor interaction $J_2$ or the long-range dipolar interaction $J_{DD}$ (12-14) and then the system orders into the classic $\sqrt{3}\times\sqrt{3}$ state at the lowest temperature.

Interesting proposals were made how the different kagome spin ice models develop into the ground state as temperature decreases. Pioneering analytical and numerical studies have established that $J_2$ vs. $J_{DD}$ lead to different symmetry-breaking pathways: A $J_1$-$J_2$ model orders through a floating Berezinskii-Kosterlitz-Thouless (BKT) critical region separated from the paramagnetic and the ground state by two BKT transitions; whereas a $J_1$-$J_{DD}$ model goes through a novel magnetic-charge-ordered (MCO) state, separated by a 3-state Potts transition and an Ising transition from the ground state and the paramagnetic state, respectively (12-14). In the MCO state, the local spins are fluctuating but the sum of magnetic charge on the two triangular plaquettes of kagome lattice form a long-range order, leaving an entropy 0.108$k_B$ per spin. The MCO state is a genuine Coulomb phase, reminiscent of the spin ice behavior on pyrochlore $(Dy\&Ho)_2Ti_2O_7$ system (28, 29).

Despite the theoretical interest, kagome spin ice states have only been realized in artificial spin ice systems (30-32), or as a metastable phase in the pyrochlore spin ice (28, 29). In these systems, symmetry-breaking pathways consistent with the $J_1$-$J_{DD}$ model have been observed. Recently, some of us demonstrated that the intermetallic HoAgGe is a faithful solid-state realization of kagome spin ice (15). The strong local easy-axis anisotropy together with ferromagnetic $J_1$ of the $Ho^{3+}$ moments lead to the ice rules on a distorted kagome lattice formed by Ho atoms in each (001) plane (15). The kagome ice rules in HoAgGe are also consistent with the various phases accessed by varying magnetic field and temperature. However, the precise nature of the zero-field intermediate phase in HoAgGe and that of the two transitions separating it from the ground and the paramagnetic states are still unclear, despite the experimental and numerical evidence suggesting a partially ordered state, which is different from any of the established scenarios mentioned above.

As another consequence of the kagome ice rules coexisting with non-negligible further-neighbor interactions, at low temperatures, the magnetization of HoAgGe versus external magnetic fields parallel to the kagome plane exhibits a series of plateaus (15-16, 22-27). More recently it was discovered that an emergent time-reversal-like degeneracy appears at the 1/3 and 2/3 plateaus as revealed by the anomalous Hall effect (AHE) (16, 33-40). However, the ground state plateau has vanishing AHE and net magnetization due to its antiferromagnetic (AFM) nature.

In this article, we use spin-polarized diffuse neutron scattering to unveil important details of the spin order in HoAgGe at different temperatures. We find that the intermediate phase is better characterized as a gradual charge-ordering crossover, following a phase transition at higher temperature in the 3D XY universality class (12-14). Moreover, magnetic and thermodynamic measurements on the $\sqrt{3}\times\sqrt{3}$ ground state reveal its unusual TRS-broken nature in terms of a finite and hysteretic nonlinear magnetic susceptibility, also originating from the ice rule. This type of TRS-breaking in the ground state of HoAgGe applies to both ideal and distorted (as in HoAgGe) kagome lattices and represents a novel prototypical nonlinear chiral AFM phase.

**Spin-polarized diffuse neutron scattering results**

The short-range ice correlations have been verified to establish in HoAgGe below 20K through thermodynamic probes (15). As a sanity check, here we first perform polarized neutron diffuse scattering measurements on co-aligned HoAgGe single crystals (see Fig. S8 for experiment setup (41)) above $T_2$ = 11.6 K at 15 K. With neutron dipole moments polarized along the *c* axis of HoAgGe crystal, aside from any nuclear contributions, the spin flip (SF) channel is related to the $M_{ab}$ components dominated by the in-plane Ising spin (41) and should therefore reveal the short-range ice correlation at this temperature. This is indeed the case as shown in Fig. 1F: The triangular shape of diffuse pattern at *K* point of the first Brillouin zone indicates the establishment of the kagome ice rule, i.e., there is one magnetic charge per triangular plaquette (see Fig. 1(c)) (17, 18). Below we name this ice-correlated paramagnetic phase kagome ice I or KI. MC simulations using the classical spin model also give similar patterns in KI (see Fig. 1(g)).

The SF diffuse scattering map at 10 K (Fig. 1(e)), below the transition temperature $T_2$ reveals the emergence of long-range order at *K* points, corresponding to a $\sqrt{3}\times\sqrt{3}$ magnetic unit cell (see Fig. 1(e) and Fig. 2(a)). In this intermediate phase denoted as kagome ice II (KII), the Ho spins are only partially ordered as seen from the persisting diffuse scattering features. Moreover, magnetic contributions to the neutron intensities at the nuclear sites are vanishingly small. This suggests that the three inequivalent ordered Ising spins in the $\sqrt{3}\times\sqrt{3}$ cell, denoted by $(\bar{\sigma}_{\mathrm{Ho1}},\bar{\sigma}_{\mathrm{Ho2}},\bar{\sigma}_{\mathrm{Ho3}})$, must add up to zero (see supplemental material for details). Previously, by assuming a partial order similar to that in (11), we performed neutron refinement with $(\bar{\sigma}_{\mathrm{Ho1}},\bar{\sigma}_{\mathrm{Ho2}},\bar{\sigma}_{\mathrm{Ho3}}) = (\bar{\sigma},-\bar{\sigma},0)$ at 10 K and obtained $\bar{\sigma}$ = 5.2(1) $\mu_B$ (Table 1 and fig. S10B in Ref. 15).

The vanishing total of $(\bar{\sigma}_{\mathrm{Ho1}},\bar{\sigma}_{\mathrm{Ho2}},\bar{\sigma}_{\mathrm{Ho3}})$ in KII also means that the ordered spins have zero net magnetic charge per triangle, distinguishing it from the ground state, which has ordered magnetic charge $\bar{Q}_m = \pm 1$ per triangle. The KII phase is therefore characterized by a "divergence-free" triad of $(\bar{\sigma}_{\mathrm{Ho1}},\bar{\sigma}_{\mathrm{Ho2}},\bar{\sigma}_{\mathrm{Ho3}})$ which breaks lattice translation symmetry, but still has fluctuating magnetic charges (19). Compared to the charge-ordered phase in dipolar kagome ice (12-14), which can be described using the above quantities as $\bar{Q}_m = \pm 1$ and $\bar{\sigma}_{\mathrm{Ho1}} = \bar{\sigma}_{\mathrm{Ho2}} = \bar{\sigma}_{\mathrm{Ho3}} = 0$, KII has $\bar{Q}_m = 0$ and $\bar{\sigma}_{\mathrm{Ho1}} + \bar{\sigma}_{\mathrm{Ho2}} + \bar{\sigma}_{\mathrm{Ho3}} = 0$, although the fluctuating spins still obey the ice rule as depicted in Fig. 1(b).

We note that $(\bar{\sigma}_{\mathrm{Ho1}},\bar{\sigma}_{\mathrm{Ho2}},\bar{\sigma}_{\mathrm{Ho3}}) = (\bar{\sigma},-\bar{\sigma},0)$ is not the only possibility that satisfies the above constraint. We have refined the elastic magnetic neutron data at 10 K using the $(\frac{\bar{\sigma}}{2}, \frac{\bar{\sigma}}{2}, -\bar{\sigma})$ and $(\frac{\bar{\sigma}}{3}, \frac{2\bar{\sigma}}{3}, -\bar{\sigma})$ states (see Fig.S1) that have the same magnetic space group and obtain almost identical refinement factors *R* and *wR* as in the previous case, but with $\bar{\sigma} = 6.0(1)\,\mu_B$ and $5.9(1)\,\mu_B$, respectively. In fact, any ordered states that fulfill the divergence-free constraint would yield virtually the same refinement factors.

As the temperature continues to decrease below 10 K, the neutron diffuse scattering intensities gradually disappear, accompanied by the increase of magnetic contributions at the nuclear sites, as shown in Fig. 1(d) and Fig. 2(a). Despite being small, the magnetic contribution at nuclear site (2, -1, 0) arises below $T_2$ (see Fig.S2(b)), while its behavior resembles a power-law spin correlation in the vicinity of $T_1 \sim 7$ K (see Fig.S2(c)). At 4 K, the system reaches its ground state $(\bar{\sigma}, \bar{\sigma}, -\bar{\sigma})$, with the ordered spin $\bar{\sigma} = 7.5(1)\,\mu_B$ (see Fig. 1(a)) (15). The evolution of $T_1$ and $T_2$ transitions under magnetic fields H//*b* is shown in the phase diagram Fig. 2(b) based on low-temperature magnetic specific heat $C_{mag}$ in Fig. S3.

**Macroscopic evidence of TRS breaking below $T_2$**

The TRS breaking nature of the ground state of HoAgGe can be captured by magnetometry as well as magnetostriction measurements under varying H//*b* around zero field. Fig. 2(c-e) summarizes field- and temperature-dependent magnetization *M*, its derivative *dM/dH*, and second derivative $d^2M/dH^2$ data (see Supplementary Fig. S4 and Fig. S5 for details). First of all, the absence of net *M* at zero field under field sweeps between ±2.25 T is consistent with the AFM nature of the ground state. However, pronounced hysteretic behavior is observed in *M*, *dM/dH*, and $d^2M/dH^2$ curves below 10 K, particularly with $d^2M/dH^2$ having finite values at zero field. As depicted in Fig.S4, the hysteretic signal merges into the phase boundary of 1/6 plateau under $H_b$ = 0.8 T at 2.4 K (see Fig.2(b) and Fig.S4(c)). The coercive field of hysteresis significantly decreases with increasing temperature, reaching $H_b$ = 0.12 T at 10 K, reminiscent of the finite magnetic contributions at nuclear site (2, -1, 0) below $T_2$ (see Fig.S2(b)). The field and temperature dependence of the hysteretic peaks in *dM/dH* curves (see Fig. S4 and Fig. S5) is summarized in Fig. 2(h).

Intriguingly, the low field hysteretic signature is also observed in the relative length change $\Delta L/L$ and its field derivative $d(\Delta L/L)/dH$, i.e. magnetostriction $\lambda_b$, under H//*b* (see Fig.2f-g and Fig.S6). Sweeping fields from +4 T to -4 T, the metamagnetic transitions between the ground state, 1/6 plateau, 1/3 plateau, 2/3 plateau, and saturated states are accompanied by large $\lambda_b$ peaks due to the first order nature of such transitions below 4 K. In particular, $\lambda_b$ becomes as large as $1.5\times10^{-4}$/T (between the ground state and the 1/6 plateau) and $2.1\times10^{-4}$/T (between the 1/3 and 2/3 plateau), comparable with that in the isostructural heavy-fermion AFM CePdAl and YbAgGe (42-43).

With increasing temperature, the change of the height and sharpness of the $\lambda_b$ peaks suggests an evolution from first- to second-order transitions. This is consistent with the magnetic specific heat $C_{mag}/T$ data with prominent peaks near the metamagnetic transitions below $T_2$ (see Fig.2b and Fig.S3), which are notably suppressed below 4K. In short, both $C_{mag}$ and $\lambda_b$ data demonstrate the second-order phase transition lines approaching a first-order one extending to low temperatures in the phase diagram of HoAgGe under H//*b* (see Fig.2(b)), which has also been observed in CePdAl and YbAgGe (42-43). The zoomed-in $\lambda_b$ versus *H* curves in Fig.2(h) give almost the same coercive fields as that from the *M(H)* and *dM/dH* curves (see Fig. 2(c-e)).

Taken together, the hysteretic behavior observed in both the magnetization and magnetostriction data provides compelling evidence for the presence of TRS breaking in the zero-field, low-temperature phases of HoAgGe.

**Monte Carlo simulation and scaling analysis**

To understand the critical behavior of HoAgGe from neutron and thermodynamic experiments, we perform extensive Monte Carlo (MC) simulations of a quasi-2D kagome spin ice model on the distorted kagome lattice similar to that in (15) but with an inter-layer exchange coupling $J_c$ (see Methods for details (41)).

Specific heat $C_{mag}$ from our MC simulations (Fig. 3(e)) clearly exhibit two features: A sharp peak appears at $T_2$ =17.8 K, with obvious system size dependence; another broad anomaly starts below 10 K and centers at $T_1$ = 6.5 K, insensitive to system sizes. The two specific heat features qualitatively agree with that in the experimental data (Fig. 2(a)). In addition, the entropy $S_{mag}$ in Fig. 3(e) decreases from 0.693 $k_B$ per spin of Ising paramagnet to 0.52 $k_B$ per spin around 20 K, indicative of the short-range ice order, consistent with the magnetic structure factor *S(Q)* data in Fig.1E. After $T_2$, $S_{mag}$ quickly decreases to 0.262 $k_B$ per spin at 10 K and eventually approaches zero after the broad anomaly in $C_{mag}$ around $T_1$.

The peak values of the $C_{mag}$ at $T_2$ show saturation as the size of the system increases. Therefore, the critical exponents $\alpha$ is expected to be a small, negative value, reminiscent of the $\alpha$ = -0.02 of 3D XY model (20, 44-50). A finite-size scaling analysis with the 3D XY critical exponents $\alpha$=-0.02 and $\nu$=0.67 indeed shows reasonable data collapse close to $T_2$ (Fig. 4(a)).

To gain a deeper understanding of why the transition at $T_2$ is of 3D XY universality, we further investigate the order parameter in the KII phase immediately below $T_2$. First consider the complex order parameter *M* as in (14, 21)

$$M = \frac{1}{N}\sum_{i=1}^{N} \sigma_i \exp\left(i\vec{Q}\cdot\vec{r_i}\right) \qquad (1)$$

with *N* the total number of spins and $\vec{Q} = \left(\frac{4\pi}{3}, 0\right)$. The six-fold degenerate ground states are captured by the discrete phase of $M = |M|e^{i\phi}$ with $\phi = \frac{n\pi}{3}$ and integer $0 \leq n \leq 5$ (14, 21), as shown in Fig. 3(a) which plots the histogram of $M$ at 4 K on the complex plane. (The phase angle $\phi$ is slightly tilted away from $\frac{n\pi}{3}$ due to the distorted kagome lattice of our model.) At 15 K, the histogram of $M$ in Fig.3(c) is distributed around a circle centered at origin, yet not evenly, distinct from the *U(1)* symmetry in the $J_1$-$J_2$ model (14, 21). From 15 K to 10 K, the circle falls into one of six maximum regions in the clock phase (see Fig.3(b)) and eventually to the 6 sharp peaks at 4 K.

The finite $|M|$ of the histogram at 15 K suggests that long-range order with nonzero ordered local spins is forming, but cannot be fully established due to the competition between many (more than 6) degenerate states, corresponding to different values of $\phi$ around the circle. Such a behavior indeed suggests an XY order parameter with weak 6-fold anisotropy.

The precise form of the anisotropy for the XY order parameter at $T<T_2$ determines which spin order, characterized by $(\bar{\sigma}_{\mathrm{Ho1}}, \bar{\sigma}_{\mathrm{Ho2}}, \bar{\sigma}_{\mathrm{Ho3}})$, is favored. In particular, in the event that the anisotropy changes between different sets of clock orientations (14, 20), a first-order phase transition may still occur. To address this issue, we introduce a triad of ordered Ising variables $(\sigma_1, \sigma_2, \sigma_3)$ (41) analogous to the experimental $(\bar{\sigma}_{\mathrm{Ho1}}, \bar{\sigma}_{\mathrm{Ho2}}, \bar{\sigma}_{\mathrm{Ho3}})$ to study the weakly anisotropic XY order below $T_2$. A benefit of $(\sigma_1, \sigma_2, \sigma_3)$ compared to $M$ is that it can be defined locally in each magnetic unit cell and can, e.g., help map out domains of favored clock states in real space.

$(\sigma_1, \sigma_2, \sigma_3)$ subject to the constraint of vanishing magnetic charge can be represented by points on a ternary plot. In this representation, the $(\bar{\sigma}, -\bar{\sigma}, 0)$ phase corresponds to the $(\frac{1}{3}, \frac{2}{3}, 0)$ point and its equivalent on the ternary plot (Fig.4(b)). Mappings of $(\frac{\bar{\sigma}}{2}, \frac{\bar{\sigma}}{2}, -\bar{\sigma})$ and $(\frac{\bar{\sigma}}{3}, \frac{2\bar{\sigma}}{3}, -\bar{\sigma})$ discussed above are illustrated in Fig. 4(c). Fig. 4(d-e) depict the ternary plots for the intermediate state at 10 K (see Fig. S15) averaged over 1000 and 5000 steps, respectively. Apparently, the anisotropy leans more towards that of $(\frac{\bar{\sigma}}{2}, \frac{\bar{\sigma}}{2}, -\bar{\sigma})$ and $(\frac{\bar{\sigma}}{3}, \frac{2\bar{\sigma}}{3}, -\bar{\sigma})$ than to $(\bar{\sigma}, -\bar{\sigma}, 0)$. The former two states are therefore more likely to describe the KII phase of HoAgGe.

Since the $U(1)$ symmetry is fully broken at $T_2$, there should not be additional second-order phase transitions below $T_2$. Moreover, since the form of 6-fold anisotropy is consistent with that of the ground state, no first-order transitions are expected below $T_2$ either. Therefore, the broad anomaly of $C_{mag}$ at $T_1$ should be a crossover, during which the remnant fluctuating spin components in the divergence channel become ordered but without symmetry change. On the one hand, this is supported by the rather weak sample size dependence of the $T_1$ specific heat peak. On the other hand, the structure factor $S(Q)$ from our MC simulations also has vanishing intensities at the nuclear sites at 15

K, but the latter gradually become finite at 10 K (Fig. S14), consistent with experimental results. The above MC results and scaling analysis provide concrete evidence supporting the 3D XY universality in our quasi-2D model as well as in HoAgGe.

**Nonlinear susceptibility of the TRS-breaking ground state**

A prominent macroscopic feature of HoAgGe's spin states below $T_2$ is its zero-field nonlinear susceptibility $\chi^{(1)}=d^2M/dH^2$ under H//*b* (the zeroth-order counterpart, $\chi^{(0)}$, stands for the ordinary susceptibility whose superscript will be omitted below). As shown in Fig. 5(c), it increases from 0.245 $\mu_B/Ho/T^2$ at 8 K to the max value 0.287 $\mu_B/Ho/T^2$ at 6 K, then gradually decreases to 0.017 $\mu_B/Ho/T^2$ at 2.4 K. The appearance of the $M(H)$ hysteresis below $T_2$ is critically connected to the zero-field $\chi^{(1)}$, since such a nonlinear susceptibility breaks the degeneracy of the two TRS partners of the ground state under a weak magnetic field, in spite of the vanishing net magnetization of both states (51-58). We note the zero-field $\chi^{(1)}$ has a similar temperature dependence as that of $d\chi/dT$ measured under 500 Oe field, the origin of which deserves future investigations.

The ground state of HoAgGe which has the magnetic space group P-6'm2' (187.212) (15) indeed allows the existence of $\chi^{(1)}$. Standard symmetry analysis in table S1 (41) yields the following nonvanishing components of $\chi^{(1)}$ that all depend on a single parameter:

$$\chi^{(1)}_{bbb} = -\chi^{(1)}_{aab} = -\chi^{(1)}_{aba} = -\chi^{(1)}_{baa} \equiv \chi^{(1)}_{b} \tag{2}$$

Figure 5(b) plots the angular dependence of $\chi^{(1)}$ along the direction of $\mathbf{H}$ rotating in the $ab$ plane with azimuthal angle $\phi$, which has a simple form according to Eq. (2): $\chi^{(1)}_{\phi} = \cos\phi(\cos^2\phi - 3\sin^2\phi)\chi^{(1)}_{b}$. In particular, the vanishing of $\chi^{(1)}_{\phi}$ at $\phi = \pm\frac{\pi}{2}$, corresponding to H//*a*, is due to the mirror plane parallel to *a*. This is consistent with experimental observations for H//*a* below 10 K (see Fig. S7) which show zero $\chi^{(1)}$ with no hysteresis.

To get a deeper understanding on the origin of $\chi^{(1)}$ in HoAgGe and how it takes different values for the two degenerate ground states, we recall that response properties of a symmetry-breaking state are determined by the latter's elementary excitations. For a kagome spin ice such as HoAgGe, its elementary excitations are individual ice-rule-permitted spin flips out of a given ground state. Consider the ground state illustrated in Fig. 5(a), one can see that, e.g., flipping spin No. 3 or No. 9 in accordance with the ice-rule increases the net spin per magnetic unit cell along *b* by +2 (in units of each Ho's spin), however, there are no ice-rule compatible spin flips that can change the net spin by -2 (see table S2). Therefore, the degeneracy between the two TRS-breaking ground states of kagome spin ice are already lifted at the single spin-flip level.

Taking all ice-rule-permitted single-spin flips into account and using the classical spin model of HoAgGe, one can arrive at a rough estimate of the low-temperature $\chi_b^{(1)} \approx 0.78\ \mu_B/Ho/T^2$ at 4 K (41), which is of the same order of magnitude as the experimental values. Moreover, the $\chi_b^{(1)}$ due to such ice-rule-permitted excitations is expected to vanish at 0 K (no excitations) as well as $T$ approaches $T_2$ from below ($\chi_b^{(1)}$ is forbidden by symmetry above $T_2$), hence consistent with the experimental observation of its non-monotonic temperature dependence.

## Discussion

Previous studies have established that the ordering pathways of kagome spin ice can be analyzed using a generalized six-state clock model (13-14, 20-21), since the $\sqrt{3}\times\sqrt{3}$ ground state has exact 6-fold degeneracy due to translation and time reversal. However, since HoAgGe is inherently 3D, it is expected to behave similarly as the 3D 6-state clock model, which orders through a single 3D XY transition from the paramagnetic state (44-45). Nonetheless, the critical behavior of HoAgGe is not completely trivial, since it is known that anisotropic terms in 3D XY models are "dangerously irrelevant", which means that they are irrelevant in the renormalization group sense in the paramagnetic state but can change the critical behavior immediately below the transition in the ordered phase (46-50).

Such a subtle behavior is indeed observed in our experiments as well as MC simulations, in that the critical exponents are difficult to be obtained by fitting the data below $T_2$. For example, the magnetic (1/3, 1/3, 0) neutron peak can be used to extract the exponent $\beta$, which for 3D XY model should be $\beta$=0.3485. Using the peak intensity data very close to $T_2$ yields $\beta$~0.342(5) in Fig. 2(a), in good agreement with the 3D XY value. However, using the data between 9 K and 11.6 K (15) leads to different $\beta$~0.321(3). The dangerously irrelevant nature of the anisotropy term may also contribute to the slow ordering of the divergence channel below $T_2$ through a crossover.

In noncollinear AFM such as $Mn_3X$ ($X$=Ir, Pt, Sn, Ge, etc.) and $Mn_3NiN$, the nontrivial TRS-breaking nature is reflected in their weak magnetization and anomalous Hall effect similar to that of ferromagnets (36-40), and they are in this sense "chiral". The ground state of HoAgGe (see Fig.1(a)), and of kagome ice in general, belongs to a different category of chiral AFM, in the sense that the above, "linear" properties as in typical ferromagnets vanish, but the two time-reversal partners still exhibit different nonlinear responses (the magnetoresistance reported in (16) is also such an example), allowing them to be distinguished and switched by external means.

## References and Notes

1. L. Balents, Spin liquids in frustrated magnets. *Nature* **464**, 199–208 (2010)
2. M. J. Harris et al. Geometrical Frustration in the Ferromagnetic Pyrochlore $Ho_2Ti_2O_7$. *Phys. Rev. Lett.* **79**, 2554 (1997)
3. A. P. Ramirez et al. Zero-point entropy in 'spin ice'. *Nature* **399**, 333–335 (1999)
4. S. T. Bramwell, M. J. P. Gingras, Spin Ice State in Frustrated Magnetic Pyrochlore Materials. *Science* **294**, 1495–1501 (2001)
5. C. Castelnovo, R. Moessner, S. L. Sondhi, Magnetic monopoles in spin ice. *Nature* **451**, 42–45 (2008)
6. L. Pauling, The Structure and Entropy of Ice and of Other Crystals with Some Randomness of Atomic Arrangement. *J. Am. Chem. Soc* **57**, 2680–2684 (1935)
7. D. J. P. Morris et al. Dirac Strings and Magnetic Monopoles in the Spin Ice $Dy_2Ti_2O_7$. *Science* **326**, 411–414 (2009)
8. T. Fennell et al. Magnetic Coulomb Phase in the Spin Ice $Ho_2Ti_2O_7$. *Science* **326**, 415–417 (2009)
9. Wolf, M. & Schotte, K. D., Ising model with competing next-nearest-neighbour interactions on the Kagome lattice. *J. Phys. A Math. Gen.* **21**, 2195-2209 (1988).
10. Takagi, T. & Mekata, M., Magnetic Ordering of Ising Spins on Kagomé Lattice with the 1st and the 2nd Neighbor Interactions. *J. Phys. Soc. Jpn.* **62**, 3943-3953 (1993).
11. A. S. Wills, R. Ballou, C. Lacroix, Model of localized highly frustrated ferromagnetism: The kagomé spin ice. *Phys. Rev. B* **66**, 144407 (2002)
12. G. Möller, R. Moessner, Magnetic multipole analysis of kagome and artificial spin-ice dipolar arrays. *Phys. Rev. B* **80**, 140409 (2009)
13. G.-W. Chern, P. Mellado, O. Tchernyshyov, Two-stage ordering of spins in dipolar spin ice on the kagome lattice. *Phys. Rev. Lett.* **106**, 207202 (2011)
14. G.-W. Chern, O. Tchernyshyov, Magnetic charge and ordering in kagome spin ice. *Phil. Trans. R. Soc. A* **370**, 5718 (2012).
15. Zhao, K. et al., Realization of the kagome spin ice state in a frustrated intermetallic compound. *Science* **367**, 1218-1223 (2020).
16. Zhao, K. et al., Discrete degeneracies distinguished by the anomalous Hall effect in a metallic kagome ice compound. *Nat. Phys.* **20**, 442–449 (2024)
17. L. Anghinolfi et al. Thermodynamic phase transitions in a frustrated magnetic metamaterial. *Nat. Commun.* **6**, 8278 (2015).

18. Canals, Benjamin et al., Fragmentation of magnetism in artificial kagome dipolar spin ice. *Nat. Commun.* **7**, 11446 (2016).
19. M. E. Brooks-Bartlett, et al. Magnetic-Moment Fragmentation and Monopole Crystallization. *Phys. Rev. X* **4**, 011007 (2014)
20. Yao Wang, Stephan Humeniuk, and Yuan Wan, Tuning the two-step melting of magnetic order in a dipolar kagome spin ice by quantum fluctuations. *Phys. Rev. B* **101**, 134414 (2020).
21. Wen-Yu Su, Feng Hu, Chen Cheng, and Nvsen Ma, Berezinskii-Kosterlitz-Thouless phase transitions in a kagome spin ice by a quantifying Monte Carlo process: Distribution of Hamming distances. *Phys. Rev. B* **108**, 134422 (2023)
22. Li, N. et al., Low-temperature transport properties of the intermetallic compound HoAgGe with a kagome spin-ice state. *Phys. Rev. B* **106**, 014416 (2022)
23. S. Roychowdhury et al. Enhancement of the anomalous Hall effect by distorting the Kagome lattice in an antiferromagnetic material. *Proc. Natl. Acad. Sci. USA* **121**, 01970 (2024)
24. H. B. Deng et al. Local Excitation of Kagome Spin Ice Magnetism Seen by Scanning Tunneling Microscopy. *Phys. Rev. Lett.* **133**, 046503 (2024)
25. Hari Bhandari et al. Tunable topological transitions in the frustrated magnet HoAgGe. *Commun. Mater.* **6**, 52 (2025).
26. Shangfei Wu et al. Lattice dynamics and spin-phonon coupling in the kagome spin ice HoAgGe. *Phys. Rev. B* **111**, 125116 (2025)
27. F. Schilberth et al. Large magnetoreflectance and optical anisotropy due to *4f* flat bands in the frustrated kagome magnet HoAgGe. *arXiv:* 2504. 10274 (2025)
28. Y. Tabata et al. Kagomé Ice State in the Dipolar Spin Ice $Dy_2Ti_2O_7$. *Phys. Rev. Lett.* **97**, 257205 (2006)
29. T. Fennell et al. Pinch points and Kasteleyn transitions in kagome ice. *Nat. Phys.* **3**, 566–572 (2007)
30. Y. Qi, T. Brintlinger, and J. Cumings, Direct observation of the ice rule in an artificial kagome spin ice. *Phys. Rev. B* **77**, 094418 (2008)
31. E. Mengotti et al. Real-space observation of emergent magnetic monopoles and associated Dirac strings in artificial kagome spin ice. *Nat. Phys.* **7**, 68–74 (2011)
32. C. Nisoli, R. Moessner, and P. Schiffer, *Colloquium*: Artificial spin ice: Designing and imaging magnetic frustration. *Rev. Mod. Phys.* **85**, 1473-1490 (2013)
33. R. Shindou and N. Nagaosa, Orbital Ferromagnetism and Anomalous Hall Effect in Antiferromagnets on the Distorted fcc Lattice. *Phys. Rev. Lett.* **87**, 116801 (2001)

34. Y. Taguchi et al. Spin Chirality, Berry Phase, and Anomalous Hall Effect in a Frustrated Ferromagnet. *Science* **291**, 2573–2576 (2001)
35. Y. Machida et al. Time-reversal symmetry breaking and spontaneous Hall effect without magnetic dipole order. *Nature* **463**, 210–213 (2010)
36. H. Chen, Q. Niu, A. H. MacDonald, Anomalous Hall Effect Arising from Noncollinear Antiferromagnetism. *Phys. Rev. Lett.* **112**, 017205 (2014)
37. S. Nakatsuji et al. Large anomalous Hall effect in a non-collinear antiferromagnet at room temperature. *Nature* **527**, 212–215 (2015).
38. Liu, Z. Q. et al. Electrical switching of the topological anomalous Hall effect in a non-collinear antiferromagnet above room temperature. *Nat. Electron.* 1, 172–177 (2018).
39. K. Zhao et al. Anomalous Hall effect in the noncollinear antiferromagnetic antiperovskite $Mn_3Ni_{1-x}Cu_xN$. *Phys. Rev. B* **100**, 045109 (2019)
40. H. Chen, Electronic chiralization as an indicator of the anomalous Hall effect in unconventional magnetic systems. *Phys. Rev. B* **106**, 024421 (2022).
41. Materials and methods, additional figures, together with experimental and theoretical text are available as supplementary materials
42. S. Lucas et al. Entropy Evolution in the Magnetic Phases of Partially Frustrated CePdAl. *Phys. Rev. Lett.* **108**, 107204 (2017).
43. G. M. Schmiedeshoff et al. Multiple regions of quantum criticality in YbAgGe. *Phys. Rev. B* **83**, 180408(R) (2011).
44. Aloysius P. Gottlob and Martin Hasenbusch, Critical behaviour of the 3D XY-model: a Monte Carlo study. *Physica A* **201**, 593-613 (1993)
45. Seiji Miyashita, Nature of the Ordered Phase and the Critical Properties of the Three Dimensional Six-State Clock Model. *J. Phys. Soc. Jpn.* **66**, 3411-3420 (1997).
46. J. Lou, A. W. Sandvik, and L. Balents, Emergence of U(1) symmetry in the 3D $XY$ model with $Zq$ anisotropy. *Phys. Rev. Lett.* **99**, 207203 (2007).
47. M. E. Fisher, Renormalization Group in Critical Phenomena and Quantum Field Theory, J. Gunton and M. S. Green, Eds., (Temple University, 1975).
48. D. R. Nelson, Coexistence-curve singularities in isotropic ferromagnets, *Phys. Rev. B* **13**, 2222 (1976).
49. D. J. Amit and L. Peliti, On dangerously irrelevant operators, *Ann. Phys.* (N.Y.) 140, 207 (1982).
50. H. Shao, W. Guo, A. W. Sandvik, Monte Carlo renormalization flows in the space of relevant and irrelevant operators: Application to three-dimensional clock models, *Phys. Rev. Lett.* **124**, 080602 (2020).
51. M. Blume, L. M. Corliss, J. M. Hastings, and E. Schiller, Observation of an antiferromagnet in an induced staggered magnetic field: Dysprosium aluminum garnet near the tricritical point, *Phys. Rev. Lett.* **32**, 544 (1974).

52. J. F. Dillon, E. Y. Chen, N. Giordano, and W. P. Wolf, Time-reversed antiferromagnetic states in dysprosium aluminum garnet, *Phys. Rev. Lett.* **33**, 98 (1974).
53. N. Giordano and W. P. Wolf, Induced staggered magnetic fields in antiferromagnets: Microscopic mechanisms, *Phys. Rev. B* 21, 2008 (1980).
54. T. C. Fujita, Y. Kozuka, M. Uchida, A. Tsukazaki, T. Arima, and M. Kawasaki, Odd-parity magnetoresistance in pyrochlore iridate thin films with broken time-reversal symmetry, *Scientific Reports* 5, 10.1038/srep09711 (2015).
55. Tian Liang, Timothy H. Hsieh, Jun J. Ishikawa, Satoru Nakatsuji, Liang Fu and N. P. Ong, Orthogonal magnetization and symmetry breaking in pyrochlore iridate $Eu_2Ir_2O_7$, *Nat. Phys.* **13**, 599–603 (2017)
56. Yilin Wang, Hongming Weng, Liang Fu and Xi Dai, Noncollinear Magnetic Structure and Multipolar Order in $Eu_2Ir_2O_7$, *Phys. Rev. Lett.* **119**, 187203 (2017).
57. E. Lhotel et al. Evidence for dynamic kagome ice. *Nat. Commun.* **9**, 3786 (2018)
58. J. Xu et al. Anisotropic exchange Hamiltonian, magnetic phase diagram, and domain inversion of $Nd_2Zr_2O_7$. *Phys. Rev. B* **99**, 144420 (2019)

## Acknowledgements

The authors would like to thank Vaclav Petříček, Oleg Tchernyshyov, Yuan Wan, Yoshi Tokiwa, Jianhui Xu, Zheng Deng, Xiancheng Wang, Jie Shen, Huifen Ren and Shaokui Su for helpful discussions and experimental support.

**Funding:** The work was supported by the National Key R&D Program of China (Grant No. 2023YFA1406003 and 2022YFA1402703), National Natural Science Foundation of China (Grants No. 12274015 and 12474139), the Beijing Natural Science Foundation (Grant No. JQ24012), and the Fundamental Research Funds for the Central Universities. The work in Augsburg was supported by the German Research Foundation (DFG) through SPP1666 (project no. 220179758), TRR80 (project no. 107745057), TRR360 (project no. 492547816) and via the Sino-German Cooperation Group on Emergent Correlated Matter. The instrument POLI at Heinz Maier-Leibnitz Zentrum (MLZ), Garching, Germany, was operated by RWTH Aachen University in cooperation with FZ Jülich (Jülich Aachen Research Alliance JARA). A portion of this work was conducted at the Synergetic Extreme Condition User Facility (SECUF).

**Author contribution:** K.Z. and P.G. proposed the experiments; K.Z., J.S.G., and X.L.C. synthesized single crystals and conducted specific heat measurements; K.Z. and N.O. conducted magnetostriction measurements; K.Z. and C.Q.J. performed magnetometry measurements; H.D. and V.H. conducted single-crystal elastic neutron scattering; H.D., T.M., Y.X.S., and M.J.G. measured single-crystal diffuse neutron scattering; N.S.M. and C.T. performed MC simulations and scaling analysis; H.C.

proposed ternary plot representation and provided nonlinear susceptibility calculations; K.Z., H.C., and P.G. wrote the manuscript with input from all authors.

**Competing interests:** The authors declare that they have no competing interests.

**Data, code, and materials availability:** All data presented in this paper, if not present in the main text or supplementary materials, are available upon reasonable request from the corresponding authors.

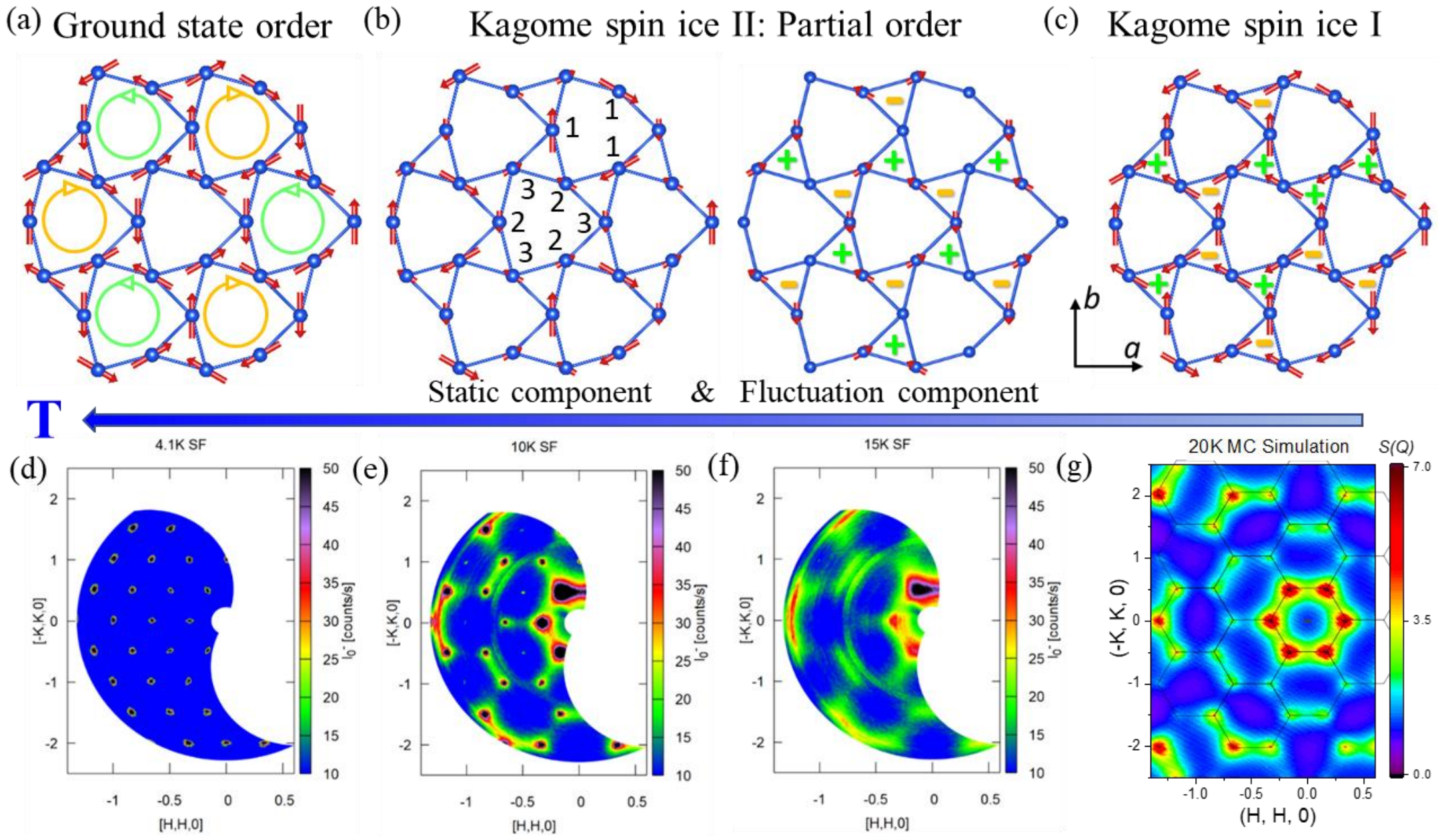

**Fig. 1 Multi-stage ordering behavior under changing temperature in kagome spin ice HoAgGe.** illustration of magnetic structure of the fully ordered ground state (a); Kagome Ice II: Partial order, including static component with ($\frac{\bar{\sigma}}{2}$, $\frac{\bar{\sigma}}{2}$, -$\bar{\sigma}$) order and related fluctuation component (b); and Kagome Ice I: short range correlation state(c), with the definition of *a* and *b* directions in (a-c) and the three inequivalent Ho sites Ho1, Ho2, and Ho3 labeled by 1, 2, and 3, respectively, for simplicity. The spin flip (SF) channel of polarized neutron diffuse scattering revealed magnetic correlations of HoAgGe in *ab* plane at 4 K (d), 10 K (e), and 15 K (f), together with structure factor *S(Q)* for the Kagome Ice I state at *T* = 20 K (g) from MC simulations based on the quasi-2D spin model (see text). The schematics of the Brillouin zones contain the vertexes of a hexagon as *K* points.

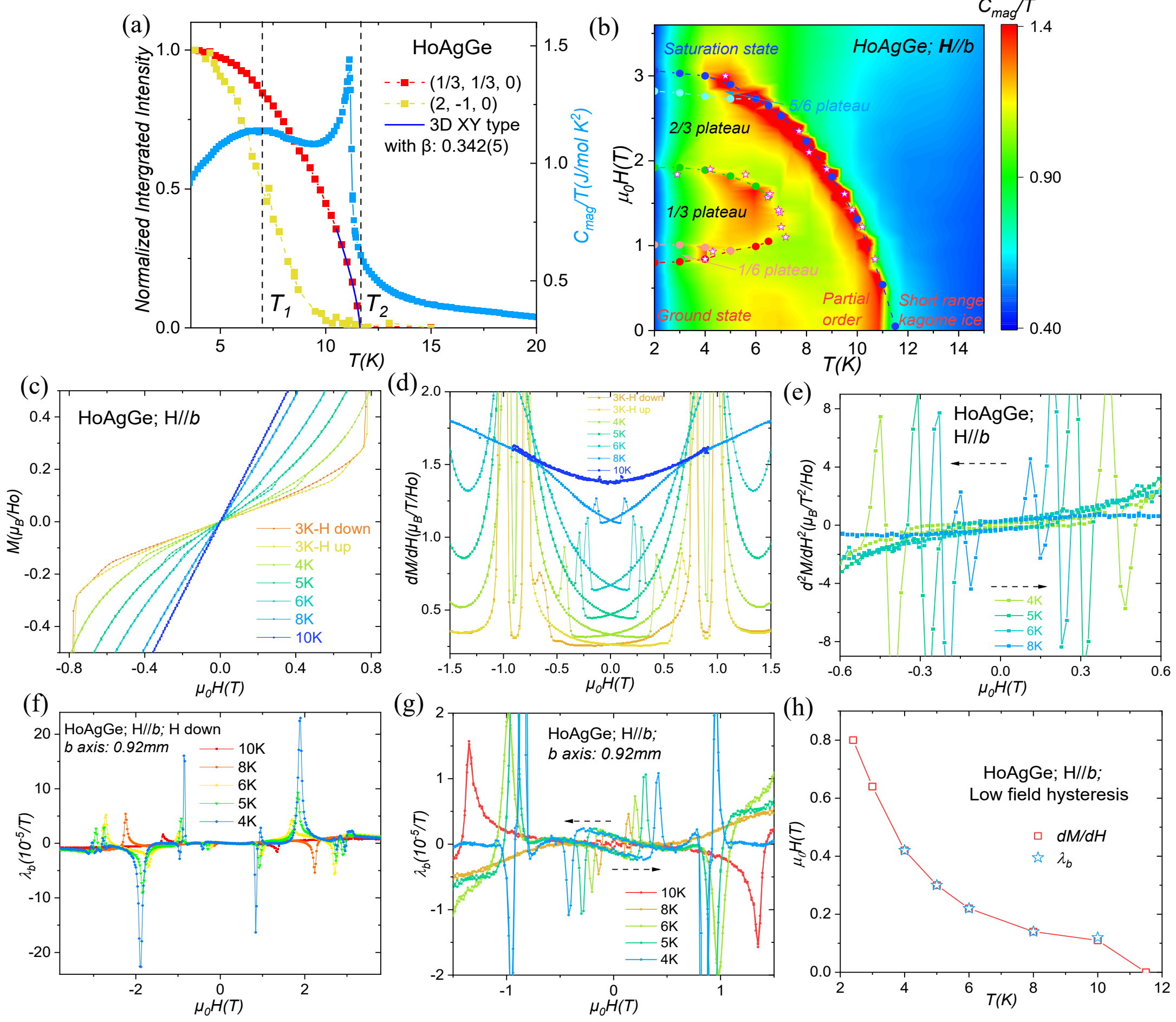

**Fig. 2. Magnetic specific heat, neutron diffraction, magnetization and magnetostriction results of HoAgGe.** (a) Integrated intensity of the magnetic peak (1/3, 1/3, 0) and nuclear site (2, -1, 0) from 15 K down to 3.8 K according to the neutron diffraction, the magnetic specific heat $C_{mag}/T$ of HoAgGe from 21K down to 2K with the dotted lines indicating onset of $T_1$ and $T_2$ (see text); (b) the H-T phase diagram of HoAgGe under H//*b* as derived from $M(H)$ (filled circles) and $C_{\mathrm{mag}}$ and $\Gamma_{\mathrm{H}}$ (empty stars) measurements, with the colour coding represents the magnetic specific heat $C_{mag}/T$ (see text). (c) The magnetization $M$, (d) its derivative $dM/dH$, and (e) its second derivative $d^2M/dH^2$ curves of HoAgGe under H//*b* with both $H$ increase and decrease condition at various temperatures; (f) the field-dependent magnetostriction $\lambda_b$ for HoAgGe along the *b* axis under H//*b* obtained upon $H$ decrease condition at various temperatures, (g) the enlargement of low field region of magnetostriction $\lambda_b$ data with both $H$ increase and decrease condition, with the increase dataset obtained through antisymmetric operation of $H$ decrease dataset in (f) on magnetic fields, (h) the field and temperature dependence of low field hysteresis for HoAgGe under H//*b* as derived from $M(H)$ (filled circles) and $\lambda_b$ (empty stars) measurements (see text).

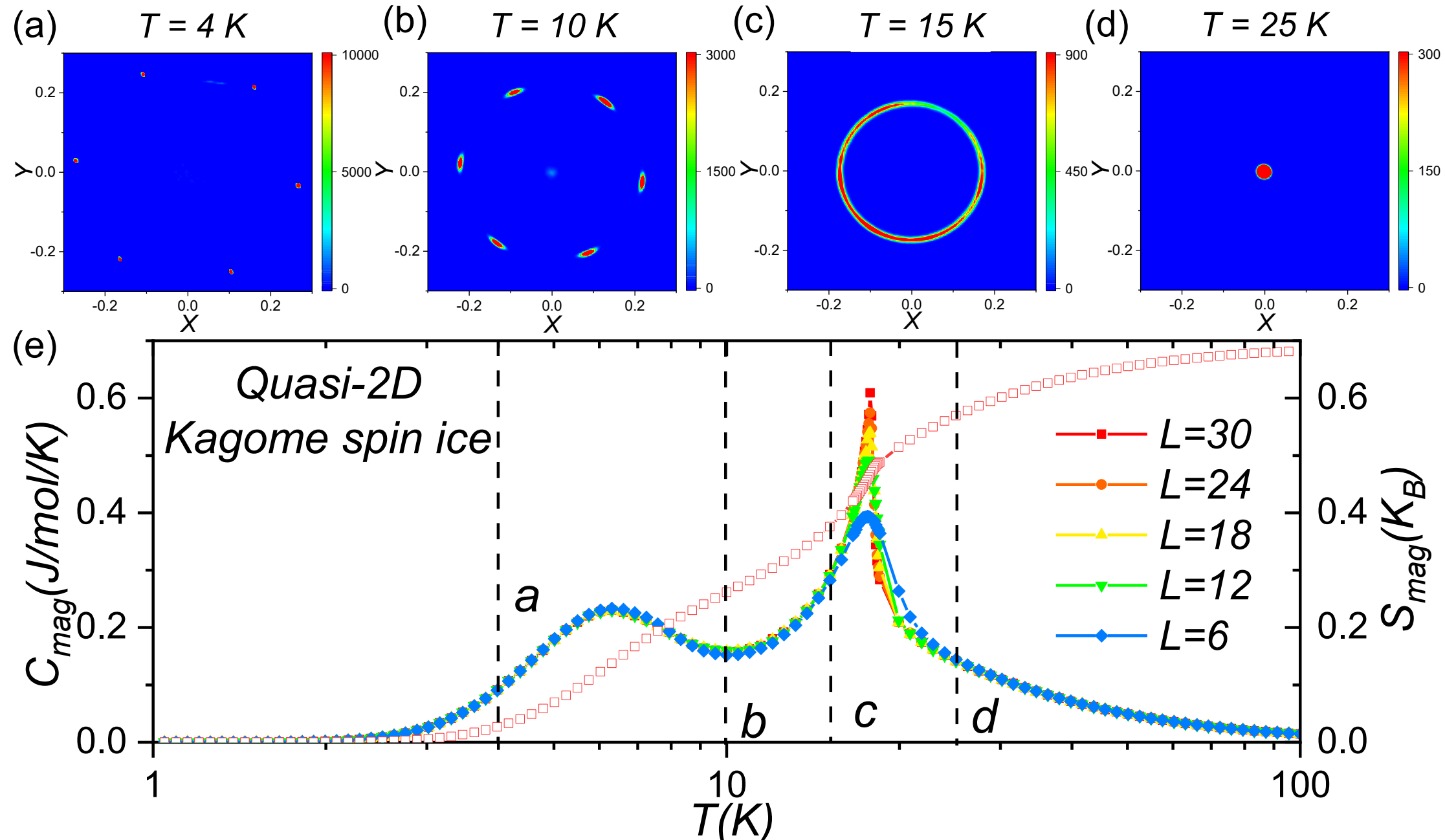

**Fig. 3 MC simulated physical property and order parameter symmetry analysis based on the spin model (see text).** Simulated low temperature magnetic specific heat $C_{mag}$ and entropy $S_{mag}$ curve with lattice size L=6, 12, 18, 24, and 30 (e). The distribution of the complex-order parameter $M$ defined in Eq. (1) at various temperature T = 4 K (a), T=10 K(b), T=15K(c), and T=25 K(d).

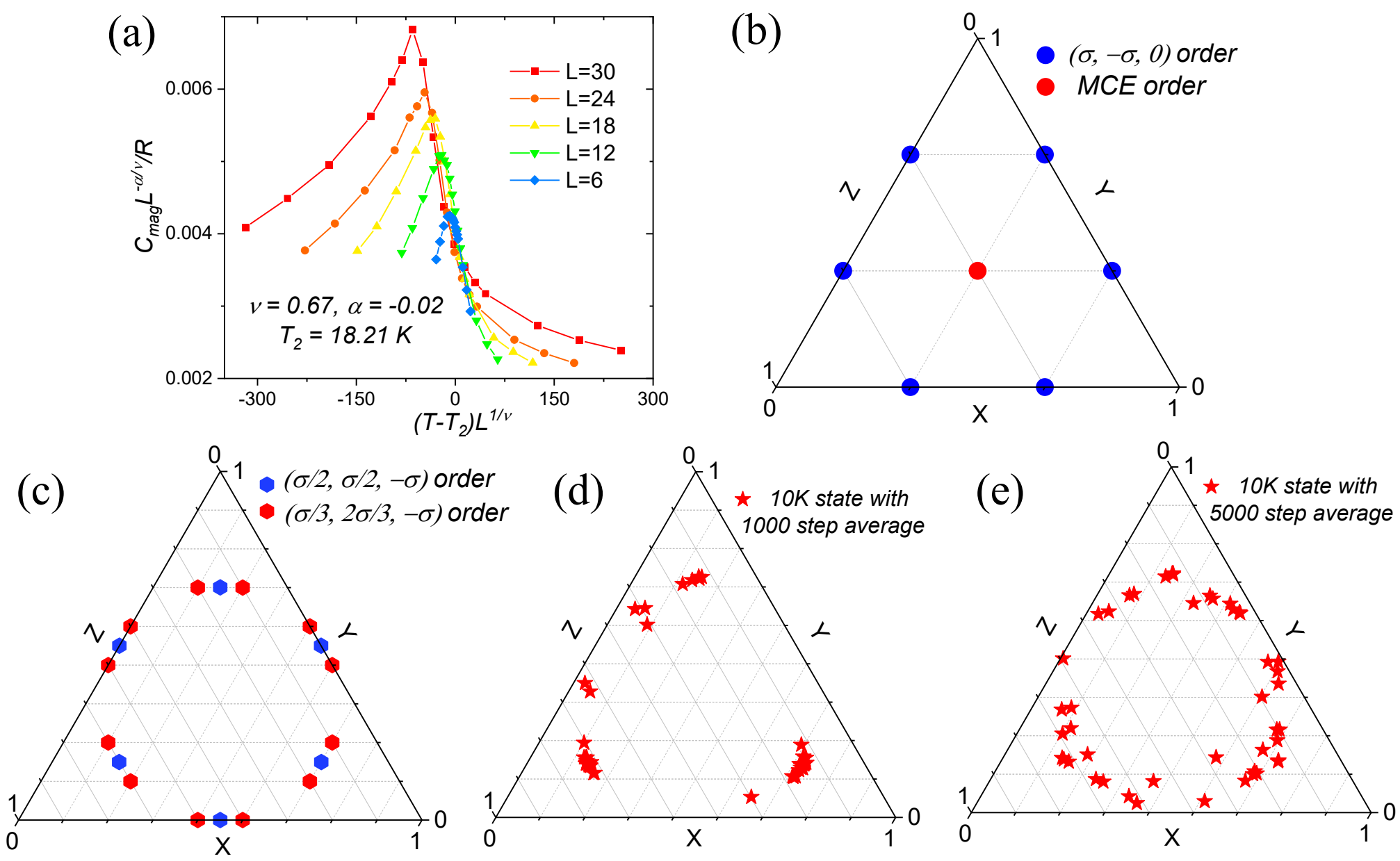

**Fig. 4 Monte Carlo simulation of the Kagome Ice II state and phase transition at *T*₂.** the scaling of Specific heat $C_{mag}$ near the $T_2$ transition using critical exponents (a) $\alpha$=-0.02 and $\nu$=0.67 expected to that of 3D XY universality class (see text), with lattice size $L$ from 6 to 30. And the ternary plots of $(\widetilde{\sigma_1},\ \widetilde{\sigma_2},\ \widetilde{\sigma_3})$ for (b) MCO phase and $(\bar{\sigma}, -\bar{\sigma}, 0)$ order; (c) $(\frac{\bar{\sigma}}{3}, \frac{2\bar{\sigma}}{3}, -\bar{\sigma})$ and $(\frac{\bar{\sigma}}{2}, \frac{\bar{\sigma}}{2}, -\bar{\sigma})$ order; the intermediate state of our kagome ice model at 10 K, with an average of (d) 1000 steps and (e) 5000 steps from the Monte Carlo simulations.

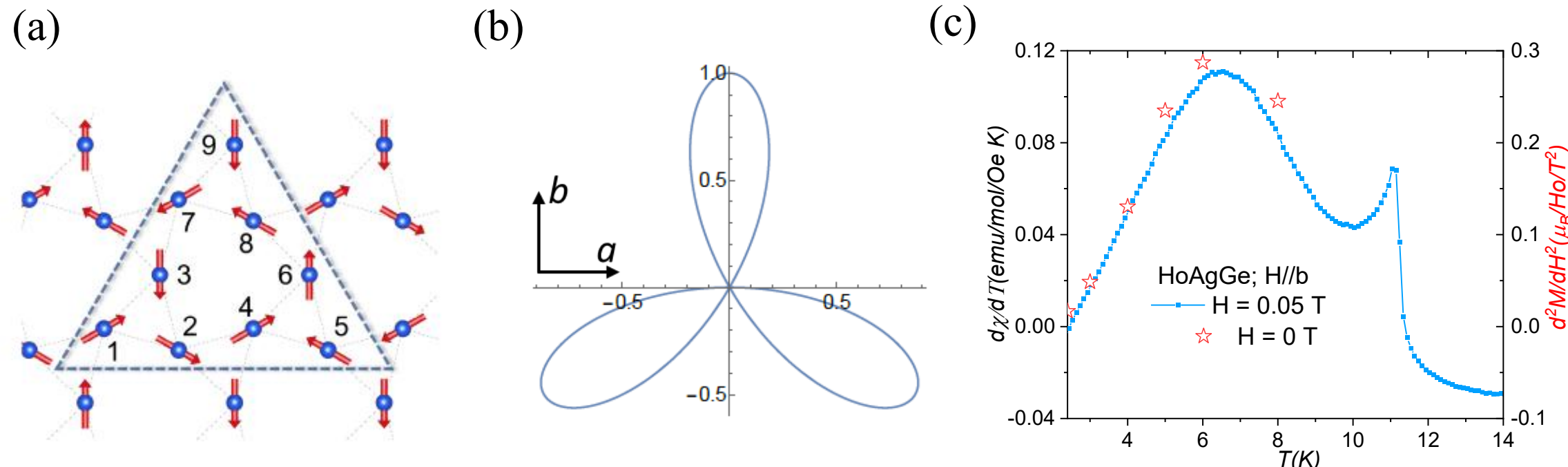

**Fig.5 Nonlinear magnetic susceptibility along *b* of HoAgGe.** (a) Ground state $S_{GS}$ of HoAgGe with only Ho spins shown. Labeled Ho spins enclosed in the dashed-line triangle are those belong to one magnetic unit cell; its TRS partner $S'_{GS}$ could be obtained through flipping all 9 Ho spins of $S_{GS}$. (b) Angular dependence of $\chi^{(1)}$ projected to the direction of a magnetic field rotating in the *ab* plane. The horizontal and vertical axes correspond to *a* and *b* axes as defined in Ref. 15. (c) Left panel: the low-temperature $d\chi/dT$ data of HoAgGe for H//*b* under 500 Oe, right panel: the $d^2M/dH^2$ value with $H = 0$ T under H//*b* at various temperatures.